\documentclass[10pt]{iopart}
\usepackage{iopams}
\usepackage{setstack}
\usepackage{amsthm}
\usepackage{graphicx}

\renewcommand{\vec}{\boldsymbol}
\renewcommand{\leq}{\leqslant}
\renewcommand{\geq}{\geqslant}
\newcommand{\bigO}{\mathcal{O}}

\eqnobysec

\newtheorem{theorem}{Theorem}[section]
\newtheorem{corollary}{Corollary}[section]
\newtheorem{prop}{Proposition}[section]

\begin{document}

\title{Late-time asymptotic dynamics of Bianchi VIII cosmologies}
\author{J T Horwood\dag, M J Hancock\ddag, D The\S~and J Wainwright\dag}
\address{\dag Department of Applied Mathematics, University of Waterloo,
  Waterloo, Ontario, Canada N2L 3G1}
\address{\ddag Department of Civil and Environmental Engineering,
  Massachusetts Institute of Technology, Cambridge, MA 02139-4307}
\address{\S Department of Mathematics, University of British Columbia,
  Vancouver, British Columbia, Canada V6T 1Z2}
\begin{abstract}
  In this paper we give, for the first time, a complete description of
  the late-time evolution of non-tilted spatially homogeneous cosmologies
  of Bianchi type VIII. The source is assumed to be a perfect fluid with
  equation of state $p = (\gamma-1)\mu$, where $\gamma$ is a constant
  which satisfies $1 \leq \gamma \leq 2$. Using the orthonormal frame
  formalism and Hubble-normalized variables, we rigorously establish the
  limiting behaviour of the models at late times, and give asymptotic
  expansions for the key physical variables.

  The main result is that asymptotic self-similarity breaking occurs,
  and is accom\-panied by the phenomenon of \textit{Weyl curvature dominance},
  characterized by the divergence of the Hubble-normalized Weyl curvature
  at late times.
\end{abstract}
\pacs{0420H, 0440N, 9880H}


\section{Introduction} \label{sec_intro}

A long term goal in theoretical cosmology is to understand the structure
and properties of the space of all cosmological solutions of the Einstein
field equations (EFEs), with a view to shedding light on the evolution of the
physical universe from the Planck time onward. In particular one wants to
study deviations from the familiar Friedmann-Lema\^{\i}tre (FL) models,
which describe a universe that is exactly homogeneous and isotropic on a
suitably large scale. In working towards this goal one makes use of a 
symmetry-based hierarchy of cosmological models of increasing complexity,
starting with the familiar FL models:
\begin{enumerate}
  \item[i)]   FL cosmologies \vspace{-0.5ex plus 0.1ex minus 0.1ex}
  \item[ii)]  non-tilted spatially homogeneous (SH) cosmologies
                \vspace{-0.5ex plus 0.1ex minus 0.1ex}
  \item[iii)] tilted SH cosmologies \vspace{-0.5ex plus 0.1ex minus 0.1ex}
  \item[iv)]  $G_{2}$ cosmologies \vspace{-0.5ex plus 0.1ex minus 0.1ex}
  \item[v)]   $G_{1}$ cosmologies \vspace{-0.5ex plus 0.1ex minus 0.1ex}
  \item[vi)]  generic cosmologies
\end{enumerate}
The terminology used in this hierarchy has the following meaning.
A SH cosmology is said to be \textit{tilted} if the fluid velocity vector is
not orthogonal to the group orbits, otherwise the model is said to be
\textit{non-tilted}. A $G_{2}$ cosmology admits a local two-parameter Abelian
group of isometries with spacelike orbits, permitting one degree of freedom
as regards spatial inhomogeneity, while a $G_{1}$ cosmology admits one
spacelike Killing vector field.

An important mathematical link between the various classes in the hierarchy is
provided by the idea of representing the physical state of a cosmological
model at an instant of time by a point in a \textit{state space}, which is
finite dimensional for classes \mbox{i)--iii)} and infinite dimensional
otherwise. The EFEs are formulated as first order evolution equations, and
\textit{the evolution of a cosmological model is represented by an orbit}
(i.e.\ a solution curve) \textit{of the evolution equations in the state
space}. The state space of a particular class in the hierarchy is a subset of
the state spaces of the more general classes, which implies that the particular
models are represented as special cases of the more general models. This
structure opens the possibility that the evolution of a model in one class
may be approximated, over some time interval, by a model in a more special
class.

Detailed information about the evolution of cosmological models more general
than FL models can only be obtained using numerical simulations or
perturbation theory. On the other hand, by introducing suitably normalized
variables, one can hope to use methods from the theory of dynamical systems to
obtain qualitative information about the \textit{asymptotic regimes} of
cosmological models, namely
\begin{enumerate}
  \item[a)] the approach to the initial singularity, characterized by
  $\ell \rightarrow 0^{+}$, and\vspace{-0.5ex plus 0.1ex minus 0.1ex}
  \item[b)] the late-time evolution, characterized by $\ell \rightarrow
  +\infty$,
\end{enumerate}
where $\ell$ is the overall length scale. From a dynamical systems point of
view, the evolution in the asymptotic regimes is governed by the dynamics on
the past attractor and the future attractor, respectively. From a physical
point of view the asymptotic regime $\ell \rightarrow 0^{+}$ corresponds to
the approach to the Planck time, while the asymptotic regime
$\ell \rightarrow +\infty$ could describe the later stages of a particular
epoch in the evolution of the universe, for example the radiation-dominated
epoch in the early universe.

The above comments provide the background for the present paper, which deals
with one of the unsolved problems concerning non-tilted SH cosmologies.
There are two generic classes of non-tilted ever-expanding SH cosmologies,
namely Bianchi type VIII and exceptional Bianchi type VI$_{-1/9}^{\ast}$,
neither of which have been fully analyzed. In this paper our interest lies
in the late-time asymptotic regime of Bianchi VIII models. Two important
results concerning this regime have been proved. Firstly, a theorem of Wald
(1983) shows that Bianchi VIII models \textit{with a cosmological constant}
are asymptotic at late times to the de~Sitter model, subject to rather weak
restrictions on the stress-energy tensor. Secondly, Ringstr\"{o}m (2001) has
recently determined rigorously the late-time behaviour of \textit{vacuum}
Bianchi VIII models\footnote[1]{Barrow and Gaspar (2001) have also studied
this problem but their results are not conclusive.}. We will comment on his
results later. On the other hand, little is known about non-vacuum models with
zero cosmological constant, apart from a brief heuristic discussion of models
containing dust or radiation by Doroshkevich \etal (1973) and by Lukash (1974).
We likewise comment on these results later. In the present paper we give a
rigorous analysis of the asymptotic dynamics at late times of Bianchi VIII
cosmologies whose matter content is a perfect fluid with equation of state
$p = (\gamma-1)\mu$, where $\gamma$ is a constant. This equation of state
includes the physically important cases of dust ($\gamma = 1$) and
radiation ($\gamma = \frac{4}{3}$).

To achieve our goal we formulate the EFEs for the Bianchi VIII cosmologies as
an asymptotically autonomous system of ordinary differential equations, using
the Hubble-normalized variables first introduced by Wainwright and Hsu (1989)
and modified by Wainwright \etal (1999)\footnote[1]{This work will henceforth
be referred to as WHU.} and Nilsson \etal (2000) in their study of the more
special Bianchi VII$_{0}$ models. We are then able to apply a theorem of
Strauss and Yorke (1967) concerning the solutions of asymptotically autonomous
systems of differential equations.

The paper is organized as follows. In \sref{sec_eveqn} we present the evolution
equations for SH cosmologies of Bianchi type VIII using the orthonormal frame
formalism and Hubble-normalized variables. The equations are derived in
detail in Wainwright and Ellis (1997)\footnote[2]{This work will henceforth
be referred to as WE.} (see chapters 5 and 6). A change of variable then leads
to a new form of the evolution equations that is adapted to the oscillatory
nature of the Bianchi VIII models. In \sref{sec_limits} we present the main
results, namely theorem~\ref{theorem_limits} and
corollary~\ref{corollary_limits}, which give the limits in the late-time
regime of the expansion-normalized variables and of certain physical
dimensionless scalars, thereby describing the asymptotic dynamics. In
\sref{sec_asympt} we strengthen the results of the theorem and corollary 
by giving the asymptotic form of the expansion-normalized variables and
physical scalars at late times. The notions of asymptotic self-similarity
breaking and Weyl curvature dominance are discussed in \sref{sec_discussion}.
Finally, in \sref{sec_overview} we conclude by giving an overview of the
asymptotic dynamics of non-tilted SH perfect fluid cosmologies, noting that
the present paper and the accompanying paper Hewitt \etal (2002) fill the gaps
that exist in the literature.

There are four appendices. Appendix~A contains the proof of the fact that
Bianchi VIII universes are not asymptotically self-similar at late times.
Appendix~B contains the proof of theorem~\ref{theorem_limits} while
in appendix~C we provide an outline of the derivation of the asymptotic
forms at late times. Finally, in appendix~D we give expressions
for the components of the Weyl curvature tensor in terms of the
expansion-normalized variables.


\section{Evolution equations} \label{sec_eveqn}

In this section we give the evolution equations for non-tilted SH
cosmologies of Bianchi type VIII. As described in WE (see pp~124--5),
we use expansion-normalized variables
\begin{equation}
  (\Sigma_{+},\Sigma_{-},N_{1},N_{2},N_{3}) , \label{eq_eveqn_physical_stateA}
\end{equation}
defined relative to a group-invariant orthonormal frame $\{\vec{e}_{a}\}$,
with $\vec{e}_{0} = \vec{u}$, the fluid 4-velocity, which is normal to
the group orbits.

The variables $\Sigma_{\pm}$ describe the shear of the fluid congruence,
and the $N_{\alpha}$, $\alpha=1,2,3$ describe the intrinsic curvature
of the group orbits. The models of Bianchi type VIII are described by
the inequality $N_{1}N_{2}N_{3} < 0$. Without loss of generality, we
assume
\begin{equation}
  N_{1} < 0, \qquad N_{2} >0, \qquad N_{3} > 0. \label{eq_eveqn_restrictionsA}
\end{equation}
It is convenient to define
\begin{equation}
  N_{+} = \case{1}{2}(N_{2}+N_{3}), \qquad
  N_{-} = \case{1}{2\sqrt{3}}(N_{2}-N_{3}), \label{eq_eveqn_Npm}
\end{equation}
and replace \eref{eq_eveqn_physical_stateA} by the state vector
\begin{equation}
  (\Sigma_{+},\Sigma_{-},N_{1},N_{+},N_{-}). \label{eq_eveqn_physical_stateB}
\end{equation}
The restrictions \eref{eq_eveqn_restrictionsA} become
\begin{equation}
  N_{1} < 0, \qquad N_{+}^{2} - 3 N_{-}^{2} > 0, \qquad
  N_{+} > 0 . \label{eq_eveqn_restrictionsB}
\end{equation}

The state variables \eref{eq_eveqn_physical_stateA} and
\eref{eq_eveqn_physical_stateB} are dimensionless, having been
normalized with the Hubble scalar\footnote[1]{On account of \eref{eq_eveqn_H},
$H$ is related to the rate of volume expansion $\Theta$ of the fluid
congruence according to $H = \frac{1}{3} \Theta$.} $H$, which is related to
the overall length scale $\ell$ by
\begin{equation}
  H = \frac{ \dot{\ell} }{ \ell }, \label{eq_eveqn_H}
\end{equation}
where the overdot denotes differentiation with respect to clock time along
the fluid congruence. The state variables depend on a dimensionless
time variable $\tau$ that is related to the length scale $\ell$ by
\begin{equation}
  \ell = \ell_{0} \rme^{\tau} , \label{eq_eveqn_ell}
\end{equation}
where $\ell_{0}$ is a constant. The dimensionless time $\tau$ is related
to the clock time $t$ by
\begin{equation}
  \frac{\rmd t}{\rmd \tau} = \frac{1}{H}, \label{eq_eveqn_t_tau}
\end{equation}
as follows from equations \eref{eq_eveqn_H} and \eref{eq_eveqn_ell}.
In formulating the evolution equations we require the deceleration parameter
$q$, defined by
\begin{equation}
  q = -\frac{\ell \ddot{\ell}}{\dot{\ell}^{2}} , \label{eq_eveqn_q_defn}
\end{equation}
and the density parameter $\Omega$, defined by
\begin{equation}
  \Omega = \frac{\mu}{3 H^{2}}. \label{eq_eveqn_Omega_defn}
\end{equation}

In terms of the variables \eref{eq_eveqn_physical_stateB}, the evolution
equations (6.9) and (6.10) in WE become
\begin{eqnarray}
  \eqalign{
  \Sigma'_{+} &= -(2-q) \Sigma_{+} - 2 N_{-}^{2} + \case{1}{3} N_{1} (
    N_{1} - N_{+} ), \\
  \Sigma'_{-} &= -(2-q) \Sigma_{-} - N_{-} ( 2N_{+} - N_{1} ), \\
  N'_{1} &= (q-4\Sigma_{+}) N_{1}, \\
  N'_{+} &= (q+2\Sigma_{+}) N_{+} + 6 \Sigma_{-} N_{-}, \\
  N'_{-} &= (q+2\Sigma_{+}) N_{-} + 2 \Sigma_{-} N_{+}, }
    \label{eq_eveqn_DE_nonpolar}
\end{eqnarray}
where
\begin{eqnarray}
  q &= 2 ( \Sigma_{+}^{2} + \Sigma_{-}^{2} ) + \case{1}{2}(3\gamma-2) \Omega ,
    \label{eq_eveqn_q_nonpolar} \\
  \Omega &= 1 - \Sigma_{+}^{2} - \Sigma_{-}^{2} - N_{-}^{2} -
    \case{1}{12} N_{1}^{2} - \left( -\case{1}{3} N_{1}N_{+} \right) ,
    \label{eq_eveqn_Omega_nonpolar}
\end{eqnarray}
and ${}'$ denotes differentiation with respect to $\tau$. For future reference
we also note the evolution equation for $\Omega$:
\begin{equation}
  \Omega ' = [2q-(3\gamma-2)] \Omega . \label{eq_eveqn_Omega_DE_nonpolar}
\end{equation}

The physical requirement $\Omega \geq 0$, in conjunction with
\eref{eq_eveqn_restrictionsB}, implies that the variables $\Sigma_{+}$,
$\Sigma_{-}$, $N_{-}$, $N_{1}$ and $N_{1} N_{+}$ are bounded, but places no
restriction on $N_{+}$ itself. In fact, it will be shown in appendix~A
(see proposition~\ref{prop_Nplus}) that if $\Omega \geq 0$ and
$\frac{2}{3} \leq \gamma \leq 2$, then for any initial conditions
\begin{equation}
  \lim_{\tau \rightarrow +\infty} N_{+} = +\infty.\label{eq_eveqn_N_plus_limit}
\end{equation}
Inspection of the equations for $\Sigma'_{+}$ and $N'_{-}$ shows that
this unboundedness of $N_{+}$ will induce rapid oscillations in $\Sigma_{-}$
and $N_{-}$ at late times, which further complicates the dynamics. A similar
situation occurs in the Bianchi VII$_{0}$ models analyzed in WHU (see p~2583).
The first step in analyzing the dynamics at late times \mbox{($\tau \rightarrow
+\infty$)} is to introduce new variables which are bounded at late
times and which enable us to isolate the oscillatory behaviour associated with
$\Sigma_{-}$ and $N_{-}$. We define
\begin{eqnarray}
  \eqalign{
  \Sigma_{-} = R \cos \psi, &\qquad N_{-} = R \sin \psi, \\
  M = \frac{1}{N_{+}}, &\qquad Z^{2} = -\frac{1}{3} N_{1} N_{+},}
    \label{eq_eveqn_cov}
\end{eqnarray}
where $R \geq 0$.

In terms of the new variables $(\Sigma_{+},R,Z,M,\psi)$, the evolution
equations \eref{eq_eveqn_DE_nonpolar} can be shown to have the following form
\begin{eqnarray}
  \eqalign{
  \Sigma'_{+} &= -(2-Q)\Sigma_{+} - R^{2} + Z^{2} + 3 M^{2} Z^{4} + 
    (1+\Sigma_{+})R^{2} \cos 2\psi, \\
  R' &= \left[ (Q+\Sigma_{+}-1) + (R^{2} - 1 -\Sigma_{+}) \cos 2 \psi - 
    \case{3}{2} M Z^{2} \sin 2 \psi \right] R, \\
  Z' &= \left[ Q - \Sigma_{+} + R^{2} \left( \cos 2\psi + \case{3}{2} 
    M \sin 2 \psi \right) \right] Z, \\
  M' &= - \left[ Q + 2 \Sigma_{+} + R^{2} \left( \cos 2\psi + 
    3 M \sin 2 \psi \right) \right] M, \\
  \psi ' &= \frac{1}{M} \left[ 2 + (1 + \Sigma_{+}) M \sin 2 \psi +
    \case{3}{2} M^{2} Z^{2} (1 - \cos 2 \psi) \right], }
   \label{eq_eveqn_full_DE}
\end{eqnarray}
where
\begin{equation}
  Q = 2 \Sigma_{+}^{2} + R^{2} + \case{1}{2}(3 \gamma-2) \Omega ,
    \label{eq_eveqn_Q}
\end{equation}
and
\begin{equation}
  \Omega = 1 - \Sigma_{+}^{2} - R^{2} - Z^{2} - \case{3}{4} M^{2} Z^{4}.
    \label{eq_eveqn_Omega}
\end{equation}
The evolution equation for $\Omega$ becomes
\begin{equation}
  \Omega ' = \left[ 2Q - (3\gamma -2) + 2 R^{2} \cos 2 \psi \right] \Omega .
    \label{eq_eveqn_Omega_DE}
\end{equation}
The restrictions \eref{eq_eveqn_restrictionsB} are equivalent to
\begin{equation}
  3 M^{2} R^{2} \sin^{2} \psi < 1, \qquad M > 0, \qquad R \geq 0, \qquad
    Z > 0 . \label{eq_eveqn_restrictionsC}
\end{equation}


\section{Limits at late times} \label{sec_limits}

In this section we present a theorem which gives the limiting behaviour as
$\tau \rightarrow +\infty$ of non-tilted Bianchi VIII cosmologies when
the equation of state parameter $\gamma$ satisfies $1 \leq \gamma \leq 2$.
As a corollary of the theorem, we obtain the limiting behaviour of certain
dimensionless scalars that describe physical properties of the models,
namely the density parameter $\Omega$, defined by \eref{eq_eveqn_Omega_defn},
the \textit{shear parameter} $\Sigma$, defined by
\begin{equation}
  \Sigma^{2} = \frac{ \sigma_{ab} \sigma^{ab} }{6H^{2}},
    \label{eq_limits_Sigma_defn}
\end{equation}
where $\sigma_{ab}$ is the rate-of-shear tensor of the fluid congruence, and
the \textit{Weyl curvature parameter} $\mathcal{W}$, defined by
\begin{equation}
  \mathcal{W}^{2} = \frac{ E_{ab} E^{ab} + H_{ab} H^{ab} }{6H^{4}},
    \label{eq_limits_Weyl_defn}
\end{equation}
where $E_{ab}$ and $H_{ab}$ are the electric and magnetic parts of the Weyl
tensor, respectively (see WE, p~19), relative to the fluid congruence.

In terms of the Hubble-normalized variables, the shear parameter is given by
\begin{equation}
  \Sigma^{2} = \Sigma_{+}^{2} + R^{2} \cos^{2} \psi,
    \label{eq_limits_Sigma_explicit}
\end{equation}
which follows from \eref{eq_eveqn_cov} in conjunction with equation (6.13)
in WE. The formula for the Weyl curvature parameter is more complicated
and is provided in appendix~D.

The main result concerning the limits of $\Sigma_{+}$, $R$, $Z$ and $M$ is
contained in the following theorem. One of the limits depends on requiring
that the model is not locally rotationally symmetric\footnote[1]{See for
example, WE p~22. We note that the LRS Bianchi VIII models are described
by the invariant subset $\Sigma_{-} = N_{-} = 0$, equivalently, $R = 0$.}
(LRS).

\begin{theorem}
  For all non-tilted SH cosmologies of Bianchi type VIII, with equation of
  state parameter $\gamma$ subject to $1 \leq \gamma \leq 2$, the
  Hubble-normalized state variables $(\Sigma_{+},R,Z,M)$ satisfy
  \begin{equation}
    \lim_{\tau \rightarrow +\infty} (\Sigma_{+},R,Z,M) =
      \left( \case{1}{2}, 0, \case{\sqrt{3}}{2}, 0 \right).
      \label{eq_limits_Hubblevars}
  \end{equation}
  If the model is not LRS, then
  \begin{equation}
    \lim_{\tau \rightarrow +\infty} \frac{M}{R} = 0 .
     \label{eq_limits_MoverR}
  \end{equation} \label{theorem_limits}
\end{theorem}

\noindent \textbf{Proof.} We first consider models which are not LRS. It follows
immediately from \eref{eq_eveqn_N_plus_limit} and \eref{eq_eveqn_cov} that
\begin{equation}
  \lim_{\tau \rightarrow +\infty} M = 0 . \label{eq_limits_M}
\end{equation}
Furthermore, since $\Sigma_{+}$ and $Z$ are bounded, it follows from the
$\psi$ evolution equation in \eref{eq_eveqn_full_DE} that
\begin{equation}
  \lim_{\tau \rightarrow +\infty} \psi = +\infty. \label{eq_limits_psi}
\end{equation}
The trigonometric functions in the DE~\eref{eq_eveqn_full_DE} thus oscillate
increasingly rapidly as $\tau \rightarrow +\infty$. In order to control these
oscillations, we define new gravitational variables $\bar{\Sigma}_{+}$,
$\bar{R}$ and $\bar{Z}$ according to
\begin{eqnarray}
  \eqalign{
  \bar{\Sigma}_{+} &= \Sigma_{+} - \case{1}{4} (1+\Sigma_{+}) R^{2}
    M \sin 2 \psi , \\
  \bar{R} &= R \left[ 1 - \case{1}{4} (R^{2}-1-\Sigma_{+})
    M \sin 2 \psi \right] , \\
  \bar{Z} &= Z \left[ 1 - \case{1}{4} R^{2} M \sin 2 \psi \right] ,
  } \label{eq_limits_cov}
\end{eqnarray}
motivated by the analysis of WHU. The evolution equations for these ``barred''
variables, which can be derived from \eref{eq_eveqn_full_DE} and
\eref{eq_limits_cov}, have the following form
\begin{eqnarray}
  \eqalign{
  \bar{\Sigma}'_{+} &= -(2-\bar{Q}) \bar{\Sigma}_{+} - \bar{R}^{2} + 
    \bar{Z}^{2} + M B_{\bar{\Sigma}_{+}} , \\
  \bar{R}' &= (\bar{Q} + \bar{\Sigma}_{+} - 1 + M B_{\bar{R}}) \bar{R} , \\
  \bar{Z}' &= (\bar{Q} - \bar{\Sigma}_{+} + M B_{\bar{Z}} ) \bar{Z} , }
    \label{eq_limits_barred_DE}
\end{eqnarray}
where
\begin{equation}
  \bar{Q} = 2 \bar{\Sigma}_{+}^{2} + \bar{R}^{2} + \case{1}{2}(3\gamma -2)
    (1-\bar{\Sigma}_{+}^{2}-\bar{R}^{2}-\bar{Z}^{2}) , \label{eq_limits_Q_bar}
\end{equation}
and the $B$ terms are bounded functions in $\bar{\Sigma}_{+}$, $\bar{R}$, 
$\bar{Z}$ and in $M$ and $\psi$ for $\tau$ sufficiently large. The essential
idea is to regard $M$ and $\psi$ as arbitrary functions of $\tau$ subject
only to \eref{eq_limits_M}. Thus, \eref{eq_limits_barred_DE} is a
non-autonomous DE for
\begin{equation}
  \bar{\vec{x}} = ( \bar{\Sigma}_{+}, \bar{R}, \bar{Z} ),
    \label{eq_limits_barred_vector_defn}
\end{equation}
of the form
\begin{equation}
  \bar{\vec{x}}' = \vec{f}(\bar{\vec{x}}) + \vec{g}(\bar{\vec{x}},\tau),
    \label{eq_limits_nonautonomous_DE}
\end{equation}
where
\begin{equation}
  \vec{g}(\bar{\vec{x}},\tau) = M(\tau) \left( B_{\bar{\Sigma}_{+}},
    \bar{R} B_{\bar{R}}, \bar{Z} B_{\bar{Z}} \right),
    \label{eq_limits_g}
\end{equation}
and the form of $\vec{f}(\bar{\vec{x}})$ can be read off from the right-hand
side of \eref{eq_limits_barred_DE}. Since
\begin{displaymath}
   \lim_{\tau \rightarrow +\infty} \vec{g}(\bar{\vec{x}},\tau) =
     \mathbf{0},
\end{displaymath}
as follows from \eref{eq_limits_M}, the DE \eref{eq_limits_barred_DE}
is \textit{asymptotically autonomous} (see for example, Strauss and
Yorke~1967). The corresponding autonomous DE is
\begin{equation}
  \hat{\vec{x}}' = \vec{f}(\hat{\vec{x}}), \label{eq_limits_autonomous_DE}
\end{equation}
where
\begin{equation}
  \hat{\vec{x}} = ( \hat{\Sigma}_{+}, \hat{R}, \hat{Z} ).
    \label{eq_limits_hatted_vector_defn}
\end{equation}

Using standard methods from the theory of dynamical systems, we first show
that
\begin{equation}
  \lim_{\tau \rightarrow +\infty} (\hat{\Sigma}_{+},\hat{R},\hat{Z}) =
    \left( \case{1}{2}, 0, \case{\sqrt{3}}{2} \right ).
    \label{eq_limits_hatted_vector}
\end{equation}
Details are provided in appendix~B.1. We then use a theorem from Strauss
and Yorke (1967) (see theorem~\ref{theorem_SY} in appendix~B) to infer that
the solutions of \eref{eq_limits_nonautonomous_DE} have the same limits
as the solutions of \eref{eq_limits_autonomous_DE}, namely
\begin{equation}
  \lim_{\tau \rightarrow +\infty} (\bar{\Sigma}_{+},\bar{R},\bar{Z}) =
    \left( \case{1}{2}, 0, \case{\sqrt{3}}{2} \right ).
    \label{eq_limits_barred_vector}
\end{equation}
Details are provided in appendix~B.2. The limit of $\vec{x} = (\Sigma_{+},R,Z)$
follows immediately from this result in conjunction with the
definitions~\eref{eq_limits_cov}. Finally, \eref{eq_limits_MoverR} is derived
in appendix~B.3, using \eref{eq_limits_Hubblevars}.

The proof of theorem~\ref{theorem_limits} for the case of LRS models
is straightforward. Since $R = 0$, the oscillatory terms in the evolution
equations~\eref{eq_eveqn_full_DE} drop out, and the variable $\psi$ becomes
irrelevant. Since \eref{eq_limits_M} still holds, the resulting system of
equations is a special case of the autonomous
DE~\eref{eq_limits_autonomous_DE}, with the result that $\Sigma_{+}$ and
$Z$ have the same limits as in the non-LRS case.
$\qquad \Box$\\[1ex plus 0.3ex minus 0.3ex]
\textit{Comment.} Our proof of theorem~\ref{theorem_limits} can be
adapted to the case of vacuum models, in which case we recover the results
of Ringstr\"{o}m (2001) (see theorem~1, p~3793).

\begin{corollary}
  For all non-tilted SH cosmologies of Bianchi type VIII, with equation
  of state parameter $\gamma$ subject to $1 \leq \gamma \leq 2$, the
  shear parameter $\Sigma$ and the density parameter $\Omega$ satisfy
  \begin{displaymath}
    \lim_{\tau \rightarrow +\infty} \Sigma = \case{1}{2}, \qquad
    \lim_{\tau \rightarrow +\infty} \Omega = 0.
  \end{displaymath}
  The Weyl curvature parameter $\mathcal{W}$ satisfies
  \begin{displaymath}
    \lim_{\tau \rightarrow +\infty} \mathcal{W} = +\infty, \quad
      \text{if the model is not LRS,}
  \end{displaymath}
  and
  \begin{displaymath}
    \lim_{\tau \rightarrow +\infty} \mathcal{W} = 0, \quad
      \text{if the model is LRS.}
  \end{displaymath} \label{corollary_limits}
\end{corollary}

\noindent \textbf{Proof.} These results follow directly from
theorem~\ref{theorem_limits} and equations \eref{eq_eveqn_Omega},
\eref{eq_limits_Sigma_explicit}, \eref{eq_app_Weyl_c} and
\eref{eq_app_Weyl_explicit}. In particular, if the model is not LRS, it follows
from \eref{eq_app_Weyl_c} and \eref{eq_app_Weyl_explicit} that since
$\Sigma_{+}$, $R$ and $Z$ are bounded, and $\lim_{\tau \rightarrow
+\infty} M/R = 0$, that
\begin{equation}
  \mathcal{W} = \frac{2R}{M}\left[1 + \bigO(M)\right],
    \label{eq_asympt_Weyl_explicit}
\end{equation}
as $\tau \rightarrow +\infty$. $\qquad \Box$

We have focussed our attention on models for which the source is a classical
fluid, i.e.~$\gamma$ lies in the range $1 \leq \gamma \leq 2$. To conclude
this section we state\footnote[1]{These results were reported earlier in
Wainwright (2000) (see p~1050).} the limiting behaviour of the scalars
$\Omega$, $\Sigma$ and $\mathcal{W}$ for values of $\gamma$ in the range
$0 < \gamma < 1$:
\begin{equation}
  \lim_{\tau \rightarrow +\infty} \Omega = \cases{
    1, &if$\quad 0 < \gamma \leq \frac{2}{3}$\\
    3(1-\gamma), &if$\quad \frac{2}{3} < \gamma < 1$}
    \label{eq_limits_Omega_nonclassical}
\end{equation}
\begin{equation}
  \lim_{\tau \rightarrow +\infty} \Sigma = \cases{
    0, &if$\quad 0 < \gamma \leq \frac{2}{3}$\\
    \case{1}{2}(3\gamma-2), &if$\quad \frac{2}{3} < \gamma < 1$}
    \label{eq_limits_Sigma_nonclassical}
\end{equation}
\begin{equation}
  \lim_{\tau \rightarrow +\infty} \mathcal{W} = \cases{
    0, &if$\quad 0 < \gamma \leq \frac{2}{3}$\\
    \case{3}{2}(3\gamma-2)(1-\gamma), &if$\quad \frac{2}{3} < \gamma <
      \frac{4}{5}$\\
    L \not= 0, &if$\quad \gamma = \frac{4}{5}$\\
    +\infty, &if$\quad \frac{4}{5} < \gamma < 1$}
    \label{eq_limits_Weyl_nonclassical}
\end{equation}
For the LRS models, we note that \eref{eq_limits_Omega_nonclassical} and
\eref{eq_limits_Sigma_nonclassical} also hold, while
\eref{eq_limits_Weyl_nonclassical} becomes
\begin{equation}
  \lim_{\tau \rightarrow +\infty} \mathcal{W} = \cases{
    0, &if$\quad 0 < \gamma \leq \frac{2}{3}$,\\
    \case{3}{2}(3\gamma-2)(1-\gamma), &if$\quad \frac{2}{3} < \gamma \leq 1$.}
      \label{eq_limits_Weyl_nonclassical_LRS}
\end{equation}

In the case $0 < \gamma < \frac{2}{3}$, these results follow from the
fact that for this range of $\gamma$, all Bianchi models are asymptotic at
late times to the flat FL model (see WE, theorem~8.2, p~174). The case
$\frac{2}{3} < \gamma < 1$ can be treated by the methods described in this
paper. Note that the models with $\frac{2}{3} < \gamma < 1$ provide a bridge
between the inflationary models\footnote[2]{Note that if $0 < \gamma < 
\frac{2}{3}$, the deceleration parameter $q$ is negative, as follows from
equation~\eref{eq_eveqn_q_nonpolar}.} with $0 < \gamma \leq \frac{2}{3}$
which isotropize, and the models with a classical fluid which do not, showing
that $\gamma = \frac{2}{3}$ is a bifurcation value. A second bifurcation
occurs at $\gamma = \frac{4}{5}$, at which value $\mathcal{W}$ makes the
transition to unbounded growth.


\section{The asymptotic solution at late times} \label{sec_asympt}

In this section we give asymptotic expansions as $\tau \rightarrow
+\infty$ for the state variables $\Sigma_{+}$, $R$, $Z$, and $M$ that
describe the class of Bianchi VIII cosmologies, and for the key dimensionless
physical scalars $\Omega$, $\Sigma$ and $\mathcal{W}$. We also give the
asymptotic expansion for the Hubble scalar $H$, and the asymptotic
relationships between clock time $t$ and the dimensionless time $\tau$.
The angular variable $\psi$ does not play a major role in the asymptotic
expansions. For brevity, we note its decay rate is governed by
\begin{displaymath}
  \psi ' = \frac{2}{M}\left[1 + \bigO(M)\right],
\end{displaymath}
as $\tau \rightarrow +\infty$, as follows from its evolution equation in
\eref{eq_eveqn_full_DE}.

Although the limiting values of the state variables as $\tau \rightarrow
+\infty$, given in theorem~\ref{theorem_limits}, are valid for all values
of the equation of state parameter $\gamma$ in the range
$1 \leq \gamma \leq 2$, it turns out that the asymptotic expansions depend
in a significant way on $\gamma$. In particular, the case $\gamma = 1$
has to be treated separately\footnote[1]{Note that in some parts of the
proof of theorem~\ref{theorem_limits}, the case $\gamma = 1$ has to be
treated separately.}. We now give the asymptotic expansions in the two cases,
$\gamma = 1$ and $1 < \gamma \leq 2$, assuming that the model is not LRS.
In the equations below, $C_{R}$, $C_{M}$, $C_{\Omega}$ and $C_{H}$ are
positive constants and $T = \frac{\ln \tau}{\tau}$.

\subsection*{State variables:}
\begin{eqnarray}
  \eqalign{
  \gamma = 1 &\qquad 1 < \gamma \leq 2 \\ \bs
  \Sigma_{+} = \case{1}{2} - \case{1}{2} \tau^{-1} \left[ 1 + \bigO(T)
    \right] &\qquad
  \Sigma_{+} = \case{1}{2} - \case{1}{4} \tau^{-1} \left[ 1 + \bigO(T)
    \right] \\
  R = C_{R} \tau^{-1} \left[ 1 + \bigO(T) \right] &\qquad
  R = \case{1}{2} \tau^{-1/2} \left[ 1 + \bigO(T) \right] \\
  Z = \case{\sqrt{3}}{2} - \case{1}{2\sqrt{3}} \tau^{-1} \left[ 1 + \bigO(T)
    \right] &\qquad
  Z = \case{\sqrt{3}}{2} + \case{1}{16\sqrt{3}} \tau^{-2} \left[ 1 + \bigO(T)
    \right] \\
  M = C_{M} \tau^{3/2} \rme^{-3/2\, \tau} \left[ 1 + \bigO(T) \right]
    &\qquad
  M = C_{M} \tau^{3/4} \rme^{-3/2\, \tau} \left[ 1 + \bigO(T) \right] }
      \label{eq_asympt_state_variables}
\end{eqnarray}

\subsection*{Density parameter:}
\begin{equation}
  \Omega = \cases{
    \tau^{-1} \left[ 1 + \bigO(T) \right], &if$\quad \gamma = 1$ \\
    C_{\Omega} \tau^{-1/2} \rme^{-3(\gamma-1)\tau}
      \left[ 1 + \bigO(T) \right], &if$\quad 1 < \gamma \leq 2$}
  \label{eq_asympt_Omega}
\end{equation}

\subsection*{Anisotropy parameters:}
\begin{equation}
  \Sigma = \cases{
    \case{1}{2} - \case{1}{2} \tau^{-1} \left[ 1 + \bigO(T) \right],
      &if$\quad \gamma = 1$ \\
    \case{1}{2} - \case{1}{4} \tau^{-1} \sin^{2} \psi \left[ 1 + \bigO(T)
    \right], &if$\quad 1 < \gamma \leq 2$}
  \label{eq_asympt_Sigma}
\end{equation}
\begin{equation}
  \mathcal{W} = \cases{
    \frac{2 C_{R}}{C_{M}} \tau^{-5/2} \rme^{3/2\, \tau}
      \left[ 1 + \bigO(T) \right], &if$\quad \gamma = 1$ \\
    \frac{1}{C_{M}}\, \tau^{-5/4} \rme^{3/2\, \tau}
      \left[ 1 + \bigO(T) \right], &if$\quad 1 < \gamma \leq 2$}
  \label{eq_asympt_Weyl}
\end{equation}

\subsection*{Hubble scalar:}
\begin{equation}
  H = \cases{
   C_{H} \tau^{1/2} \rme^{-3/2\, \tau}\left[1 + \bigO(T)\right],
     &if$\quad \gamma = 1$\\
   C_{H} \tau^{1/4} \rme^{-3/2\, \tau}\left[1 + \bigO(T)\right],
     &if$\quad 1 < \gamma \leq 2$}
 \label{eq_asympt_Hubble}
\end{equation}

\subsection*{Clock time:}
\begin{equation}
  t = \cases{
    \frac{2}{3 C_{H}} \tau^{-1/2} \rme^{3/2\, \tau}
      \left[1 + \bigO(T)\right], &if$\quad \gamma = 1$\\
    \frac{2}{3 C_{H}} \tau^{-1/4} \rme^{3/2\, \tau}
      \left[1 + \bigO(T)\right], &if$\quad 1 < \gamma \leq 2.$}
  \label{eq_asympt_clock_time}
\end{equation}

Our derivation of the above asymptotic expansions is given in appendix~C.
While not fully rigorous, it is nevertheless quite convincing. We first
derive the expansions for the solutions $\hat{\vec{x}}(\tau) =
(\hat{\Sigma}_{+},\hat{R},\hat{Z})$ of the autonomous
DE~\eref{eq_limits_autonomous_DE}, using centre manifold theory.
Secondly, it follows from the proof of theorem~\ref{theorem_limits} that
$M$ decays to zero exponentially. This fact enables us to verify that the
expansions are compatible with the exact evolution
equations~\eref{eq_limits_barred_DE}, in the sense that terms cancel at the
appropriate orders.

The derivation of the expansion for $\Omega$ depends on $\gamma$. In the case
$\gamma = 1$, the expansion follows immediately from \eref{eq_eveqn_Omega}.
The case $1 < \gamma \leq 2$ is more complicated and is treated in
appendix~C.1. Once $\Omega$ is known, the expansion for $H$ can be obtained
directly from equation~\eref{eq_eveqn_Omega_defn}, since it follows from the
contracted Bianchi identities that the matter density $\mu$ is given by
$\mu = \mu_{0}\, \rme^{-3\gamma\tau}$ (see for example, WE, equation~(1.99)).
Finally, knowing $H$, the expansion for the clock time $t$ can be obtained
from equation~\eref{eq_eveqn_t_tau}.

\subsection*{LRS models}

We note that the asymptotic expansions
\eref{eq_asympt_state_variables}--\eref{eq_asympt_clock_time} were derived
subject to the assumption that the model is not LRS, i.e.\ that the variable
$R$ is not identically zero. A detailed analysis shows that results for LRS
models with $\gamma = 1$ can be obtained by setting $C_{R} = 0$ in equations
\eref{eq_asympt_state_variables}--\eref{eq_asympt_Sigma} and
\eref{eq_asympt_Hubble}--\eref{eq_asympt_clock_time}. The expansion for
the Weyl curvature parameter $\mathcal{W}$ becomes
\begin{equation}
  \mathcal{W} = \case{1}{2}\tau^{-1}[1 + \bigO(T)],
    \label{eq_asympt_Weyl_LRS}
\end{equation}
For $1 < \gamma \leq 2$, the asymptotic expansions do not specialize to the
LRS case $R = 0$. A detailed analysis shows that the variables approach their
limiting values at an exponential rate. Specifically,
\begin{eqnarray}
  \eqalign{
  \Sigma_{+} &= \case{1}{2} + \bigO(\rme^{-\lambda_{1}\tau}) , \\
  Z &= \case{\sqrt{3}}{2} + \bigO(\rme^{-\lambda_{2}\tau}) , \\
  M &= C_{M}\, \rme^{-3/2\, \tau}[1 + \bigO(\rme^{-\lambda_{3}\tau})] , \\
  \Omega &= C_{\Omega}\, \rme^{-3(\gamma-1)\tau}[1 +
    \bigO(\rme^{-\lambda_{3}\tau})] ,} \label{eq_asympt_state_variables_LRS}
\end{eqnarray}
where $\lambda_{1}$, $\lambda_{2}$ and $\lambda_{3}$ are positive constants.
It follows that $\mathcal{W}$ tends to zero exponentially. In conclusion, the
main difference is that the LRS models do not exhibit Weyl curvature dominance.


\section{Discussion} \label{sec_discussion}

The most significant feature of the late-time asymptotic regime of non-tilted
SH perfect fluid cosmologies of Bianchi type VIII is that \textit{the models
are not asymptotically self-similar}, or in other words, the evolution of the
models at late times is not approximated by a self-similar model. We can draw
this conclusion because the orbits that describe the models in the
Hubble-normalized state space escape to infinity, and hence do not approach an
equilibrium point (see Wainwright~2000, p~1044; note that equilibrium points
in the Hubble-normalized state space correspond to self-similar Bianchi
cosmologies). The Bianchi VIII models share this feature with the Bianchi
VII$_{0}$ models (see WHU). However, the Bianchi VIII models
differ from the Bianchi VII$_{0}$ models in two important ways. Firstly, the
Bianchi VIII models with $1 \leq \gamma \leq 2$ are
\textit{vacuum-dominated at late times}, that is, the density parameter
$\Omega$ satisfies
\begin{displaymath}
  \lim_{\tau \rightarrow +\infty} \Omega = 0
\end{displaymath}
(see corollary~\ref{corollary_limits}). Secondly, the \textit{expansion is
anisotropic at late times} in the sense that the shear parameter $\Sigma$
satisfies
\begin{displaymath}
  \lim_{\tau \rightarrow +\infty} \Sigma = \case{1}{2}
\end{displaymath}
(see corollary~\ref{corollary_limits}). On the other hand, for the
Bianchi VII$_{0}$ models, the shear parameter $\Sigma$ tends to zero at late
times whenever the equation of state parameter satisfies
$\gamma \leq \frac{4}{3}$ (see WHU, theorem~2.3).

The breaking of asymptotic self-similarity in Bianchi VIII and Bianchi
VII$_{0}$ cosmologies manifests itself in the behaviour of the Weyl curvature
tensor, which describes the intrinsic anisotropy in the gravitational field.
Both classes of models exhibit what has been referred to as \textit{Weyl
curvature dominance}\footnote[1]{This notion has recently been incorporated
into a classification of the late-time dynamics of SH models by Barrow and
Hervik (2002), who refer to it as extreme Weyl dominance (see section~5).}
(see WHU, p~2588) which refers to the fact that Hubble-normalized scalars
constructed from the Weyl tensor are unbounded.
For Bianchi VIII models with $1 \leq \gamma \leq 2$, Weyl curvature
dominance manifests itself through the unbounded growth of the Weyl
parameter $\mathcal{W}$ as $\tau \rightarrow +\infty$
(see corollary~\ref{corollary_limits}), as is also the case for the
Bianchi VII$_{0}$ models with $1 < \gamma < 2$ (see WHU, theorem~2.4 and
equation~(3.40) noting that $\beta = \frac{1}{2}(4-3\gamma)$). On the other
hand, for Bianchi VII$_{0}$ models with $\gamma = 1$ the Weyl parameter
$\mathcal{W}$ approaches a finite limit and the unboundedness occurs in the
Hubble-normalized derivatives of the Weyl tensor.

We summarize the rate of growth of $\mathcal{W}$ below, expressed in
clock time $t$, as $t \rightarrow +\infty$. The results for Bianchi
VII$_{0}$ models follows from equations (3.26), (3.34) and (3.40) in
WHU. For Bianchi VII$_{0}$ models,
\begin{equation}
  \mathcal{W} \sim \cases{
    L \not= 0, &if$\quad \gamma = 1$,\\
    t^{2(\gamma-1)/\gamma}, &if$\quad 1 < \gamma < \frac{4}{3}$,\\
    \frac{\sqrt{t}}{(\ln t)^{3/2}}, &if$\quad \gamma = \frac{4}{3}$,\\
    t^{(2-\gamma)/\gamma}, &if$\quad \frac{4}{3} < \gamma < 2$,}
  \label{eq_discussion_Weyl_BVII0}
\end{equation}
while for Bianchi VIII models,
\begin{equation}
  \mathcal{W} \sim \cases{
    \frac{t}{(\ln t)^{2}}, &if$\quad \gamma = 1$\\
    \frac{t}{\ln t}, &if$\quad 1 < \gamma \leq 2$.}
  \label{eq_discussion_Weyl_BVIII}
\end{equation}
We note that rate of growth is largest for Bianchi VIII models with $\gamma$
in the range $1 < \gamma \leq 2$, and is independent of $\gamma$ in this
range. For Bianchi VII$_{0}$ models, the maximum rate of growth occurs
for the radiation equation of state, but is less than that for the Bianchi
VIII models.

We now comment on the relation between the late-time regime of perfect fluid
Bianchi VIII models and the corresponding vacuum models, as studied by
Ringstr\"{o}m (2001). It follows from corollary~\ref{corollary_limits} and
equation~\eref{eq_limits_Omega_nonclassical} that the value $\gamma = 1$ is
a bifurcation value governing the transition to a vacuum-dominated late-time 
regime for values $\gamma > 1$. This bifurcation manifests itself in the rate
at which the density parameter $\Omega$ tends to zero. If $\gamma = 1$,
$\Omega$ decays to zero at a power law rate while if $1 < \gamma \leq 2$, 
$\Omega$ decays exponentially (in terms of $\tau$; see
equation~\eref{eq_asympt_Omega}). In other words, if $1 < \gamma \leq 2$,
orbits approach the vacuum boundary $\Omega = 0$ in state space exponentially
fast, so that the leading asymptotic behaviour of the perfect fluid models,
as given in
equations~\eref{eq_asympt_state_variables}--\eref{eq_asympt_clock_time},
coincides with that of the vacuum models.

We conclude this section by providing a link between our results and the
earlier analysis of Doroshkevich \etal (1973) (see p~742) and Lukash (1974)
(see p~168), referred to in the introduction. In these papers the authors
give an approximate form of the line-element at late times for Bianchi VIII
models containing dust or radiation and of the energy density of the source.
In particular, the energy density is
\begin{equation}
  \mu \sim \cases{
    \frac{1}{t^{2}\ln t}, &if$\quad \gamma = 1$\\
    \frac{1}{t^{8/3} (\ln t)^{2/3}}, &if$\quad \gamma = \frac{4}{3}.$}
  \label{eq_disc_mu}
\end{equation}
where $t$ is clock time. Our asymptotic expansions for $\Omega$, $H$ and $t$
in equations~\eref{eq_asympt_Omega}, \eref{eq_asympt_Hubble} and
\eref{eq_asympt_clock_time}, in conjunction with \eref{eq_eveqn_Omega_defn},
lead to the asymptotic form for $\mu$ which agrees with \eref{eq_disc_mu}.


\section{Overview} \label{sec_overview}

In this concluding section we use the results of this paper and of the
accompanying paper Hewitt \etal (2002) to give an overview of the dynamics in
the asymptotic regimes of non-tilted SH perfect fluid cosmologies. These
cosmological models can be grouped into three main subclasses, following
Ellis and MacCallum (1969):
\begin{enumerate}
  \item[i)] class A models (Bianchi type I, II, VI$_{0}$, VII$_{0}$, VIII
    and IX),
  \item[ii)] non-exceptional class B models (Bianchi type IV, V, VI$_{h}$,
    and VII$_{h}$),
  \item[iii)] exceptional class B models (Bianchi type VI$_{h}$ with
    $h = -\frac{1}{9}$, denoted VI$_{-1/9}^{\ast}$),
\end{enumerate}
We refer to WE (pp~36--7, 41--2) for a summary of this classification.

\begin{figure}
  \begin{indented} \item[] \includegraphics[scale=0.75]{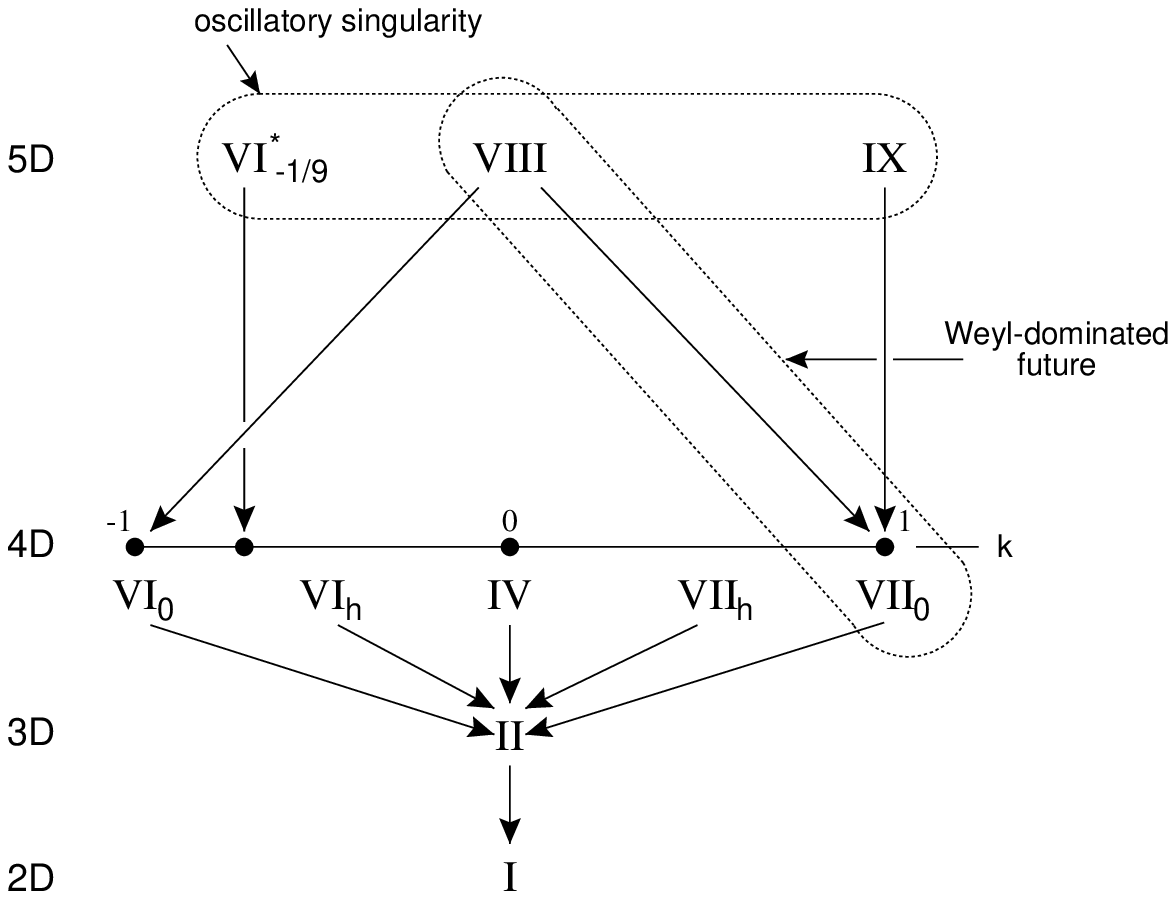}
    \end{indented}
  \caption{\label{fig_SHcosmo}Non-tilted perfect fluid spatially homogeneous
    cosmologies\footnotemark[1]. The parameter $k$ is related to the group
    parameter $h$ through $k = \tanh \frac{1}{h}$.}
\end{figure}

The dimensions of the Hubble-normalized state space for each Bianchi type
are shown in figure~\ref{fig_SHcosmo}. We note that the dimension gives
the number of arbitrary parameters in the corresponding family of solutions
(see Wainwright and Hsu (1989), p~1419).

\begin{table}
  \caption{\label{table_references}References for the asymptotic dynamics of
    non-tilted SH cosmologies with perfect fluid\footnotemark[1] source and
    equation of state $p = (\gamma-1)\mu$.}
  \begin{indented} \item[] \begin{tabular}{@{}lll} \br
  & singular regime & late-time regime \\ \mr
  class A, type I, II, VI$_{0}$ & WE, ch~5 & WE, ch~5 \\
  class A, type VII$_{0}$ & Ringstr\"{o}m (2000) & WHU and Nilsson
    \etal (2000) \\
  class A, type VIII & WE, ch~5 & present paper \\
  class A, type IX & Ringstr\"{o}m (2000) & not applicable \\
  non-exceptional class B & WE, ch~6 & WE, ch~6 \\
  exceptional class B & Hewitt \etal (2002) & Hewitt \etal (2002) \\ \br
  \end{tabular}\end{indented}
\end{table}\footnotetext[1]{For a complete description of the late-time regime
for vacuum class A models, we refer to Ringstr\"{o}m (2001).}

Table~\ref{table_references} gives the references which contain the most
comprehensive descriptions of the various classes. All of these papers make
use of the so-called Hubble-normalized variables within the framework of the
orthonormal frame formalism, and apply techniques from the theory of
dynamical systems. This approach highlights the role played by self-similar
solutions. These works in turn refer to related papers, describe other methods,
specifically the Hamiltonian approach and the metric approach (see WE,
pp~5--6), and give information about the historical development.

The table shows that with the appearance of the present paper and the paper
Hewitt \etal (2002), there is now available a complete description of the
dynamics of non-tilted SH cosmologies with perfect fluid source, in the two
asymptotic regimes. It should be noted, however, that some of the results are
based on heuristic arguments and numerical experiments. We shall mention below
what remains to be proved.

\footnotetext[1]{The Bianchi V models, for which the state space
is two-dimensional, do not fit in a natural way into this diagram.}

The principal features of the asymptotic dynamics are as follows. Firstly,
as regards the singular asymptotic regime, the three generic classes, namely
Bianchi type VIII, IX and VI$_{-1/9}^{\ast}$ (see~\fref{fig_SHcosmo}) have
an oscillatory singularity, and the asymptotic behaviour is described by a
two-dimensional attractor containing the Kasner equilibrium points in the
Hubble-normalized state space. We note that the existence of this attractor
has only been proved for the Bianchi IX models (see Ringstr\"{o}m (2000)). For
non-generic models the dynamics near the singularity is remarkably simple in
the sense that the models are asymptotically self-similar, being approximated
by a Kasner solution.

Secondly, as regards the late-time regime, the dynamics depends crucially on
whether or not the Hubble-normalized state space is bounded. For the classes
VII$_{0}$ and VIII the state space is unbounded, and the solutions exhibit
Weyl curvature dominance (see \sref{sec_discussion}). For the remaining
classes, the state space is bounded and the models are asymptotically
self-similar. In particular, all class B models, including the generic class
VI$_{-1/9}^{\ast}$, are asymptotically self-similar, although a complete
proof of this fact, except in the vacuum subcase, has not been given, due to
a lack of success in finding a monotone function for the evolution equations.

Referring to the hierarchy in the introduction, we conclude by giving some
suggestions for future research. Firstly, much work remains to be done before
the asymptotic dynamics of \textit{tilted} SH models are fully understood.
While various formulations of the EFEs have been given (see WE, p~175) the
only tilted models that have been analyzed in detail is the full class of
Bianchi II models (see Hewitt \etal 2001) and a special class of Bianchi V
models (see Hewitt and Wainwright~1992). The difficultly in analyzing the
general class of tilted SH models is highlighted by the fact that at this time
not all of the self-similar solutions have been found. Secondly, it is a
long-standing conjecture that the occurrence of oscillatory singularities in
SH models of Bianchi types VIII and IX implies that oscillatory singularities
will occur in generic spatially inhomogeneous models. Over the past five
years, considerable numerical and analytical evidence has been provided to
support this conjecture (see Weaver \etal 1998, Berger and Moncrief 1998 and
Berger \etal 2001). Likewise, the occurence of Weyl curvature dominance in SH
models of Bianchi types VII$_{0}$ and VIII leads us to conjecture that this
behaviour will also occur in spatially inhomogeneous models, perhaps as
generic behaviour. The simplest class in which this behaviour could occur is
the class of $G_{2}$ cosmologies, since the Bianchi VII$_{0}$ models are
contained as a subclass. It would thus be of interest to investigate the
$G_{2}$ cosmologies from this point of view.


\ack

The authors wish to thank Conrad Hewitt, Ulf Nilsson and David Siegel for
helpful comments and discussions. This research was supported financially by
the Natural Sciences and Engineering Research Council of Canada through a
research grant (JW) and Undergraduate Research Awards to JH (in 2002), to MH
(in 1997) and to DT (in 2000).


\appendix
\def\thesection{\Alph{section}}

\stepcounter{section}
\section*{Appendix~\thesection.~The limit of $\boldsymbol{N_{+}}$ at late
  times}

In this appendix we prove \eref{eq_eveqn_N_plus_limit}, concerning
the limit of the Hubble-normalized variable $N_{+}$ in the late-time regime.
This result is restated as proposition~\ref{prop_Nplus} below.

\begin{prop}
  For all non-tilted SH cosmologies of Bianchi type VIII, with equation
  of state parameter $\gamma$ subject to $\frac{2}{3} \leq \gamma \leq 2$ and
  density parameter $\Omega$ satisfying $\Omega \geq 0$, the Hubble-normalized
  variable $N_{+}$ satisfies
  \begin{equation}
    \lim_{\tau \rightarrow +\infty} N_{+} = +\infty .
      \label{eq_app_Nplus_limit}
  \end{equation} \label{prop_Nplus}
\end{prop}

\noindent{\textbf{Proof.}} The proof relies on concepts from dynamical systems
theory, in particular, the notion of $\omega$-limit sets (see for example,
WE, p~99) and the monotonicity principle (WE, theorem 4.12, p~103).

The Bianchi VIII state space $S$ is defined by the inequalities
\eref{eq_eveqn_restrictionsA} and $\Omega \geq 0$. The function
\begin{equation}
  \chi = ( N_{1} N_{2} N_{3} )^{2}, \label{eq_app_Nplus_chi}
\end{equation}
is positive on $S$ and satisfies
\begin{equation}
  \chi ' = 6q \chi , \label{eq_app_Nplus_chi_derivative}
\end{equation}
as follows from \eref{eq_eveqn_Npm} and \eref{eq_eveqn_DE_nonpolar}, where
$q$ is given by \eref{eq_eveqn_q_nonpolar}. If $\frac{2}{3} < \gamma \leq 2$
and $\Omega > 0$, $q$ is positive and hence $\chi$ is increasing along orbits
in $S$. On the other hand, if $\Omega = 0$ or $\gamma = \frac{2}{3}$, then
$q$ may equal zero if $\Sigma_{+} = \Sigma_{-} = 0$. Since $\Sigma_{+} =
\Sigma_{-} = 0$ is not an invariant set of the evolution
equations~\eref{eq_eveqn_DE_nonpolar}, $q$ cannot remain zero and hence
$\chi$ is again increasing along orbits in $S$.

In order to apply the monotonicity principle, we consider the set
$\bar{S} \setminus S$, where $\bar{S}$ is the closure of $S$. From the
definition of $S$, it follows that $\bar{S} \setminus S$ is defined by the
following restrictions:
\begin{equation}
  N_{1} N_{2} N_{3} = 0, \qquad N_{1} \leq 0, \qquad N_{2} \geq 0, \qquad
  N_{3} \geq 0. \label{eq_app_Nplus_state_space_boundary}
\end{equation}
By \eref{eq_app_Nplus_chi}, $\chi$ is defined and equal to zero on
$\bar{S} \setminus S$. The monotonicity principle implies that for any point
$\vec{x} \in S$, the $\omega$-limit set $\omega(\vec{x})$ is contained in the
subset of $\bar{S} \setminus S$ that satisfies the condition
$\lim_{\vec{y} \rightarrow \vec{s}} \chi(\vec{y}) \not= 0$, where
$\vec{s} \in \bar{S} \setminus S$ and $\vec{y} \in S$. This subset is the
empty set, since $\chi = 0$ on $\bar{S} \setminus S$. Therefore we conclude
that $\omega(\vec{x}) = \phi$ for all $\vec{x} \in S$.

Finally, suppose that \eref{eq_app_Nplus_limit} does not hold for each orbit
in $S$. Then there exists a number $b > 0$ such that for all $\tau_{0}$,
there exists a $\tau > \tau_{0}$ such that $N_{+}(\tau) < b$. Since the other
variables are bounded, then the orbit $\vec{x}(\tau)$ lies in a compact set
$S \subset \mathbb{R}^{5}$ and hence has a limit point in $S$, contradicting
$\omega(\vec{x}) = \phi$. Therefore \eref{eq_app_Nplus_limit} holds.
$\qquad \Box$


\stepcounter{section}
\section*{Appendix~\thesection.~Details of the proof of
  theorem~\ref{theorem_limits}}

The proof of theorem~\ref{theorem_limits} is based on a result of Strauss and
Yorke (1967) (see corollary 3.3, p~180) concerning asymptotically autonomous
DEs, stated as theorem~\ref{theorem_SY} below.

Consider a non-autonomous DE
\begin{equation}
  \bar{\vec{x}}' = \vec{f}(\bar{\vec{x}}) + \vec{g}(\bar{\vec{x}},\tau),
    \label{eq_app_SY_nonautonomous_DE}
\end{equation}
and the associated autonomous DE
\begin{equation}
  \hat{\vec{x}}' = \vec{f}(\hat{\vec{x}}), \label{eq_app_SY_autonomousDE}
\end{equation}
where $\vec{f} : D \rightarrow \mathbb{R}^{n}$, $\vec{g} : D \times
\mathbb{R} \rightarrow \mathbb{R}^{n}$ and $D$ is an open subset of
$\mathbb{R}^{n}$. It is assumed that\\[1ex plus 0.3ex minus 0.3ex]
\begin{tabular}{@{}lll}
  & $H_{1}:$ & ${\dsty \lim_{\tau \rightarrow +\infty}
    \vec{g}(\vec{w}(\tau),\tau) = \mathbf{0}} \text{ for every
    continuous function $\vec{w} : [\tau_{0},+\infty) \rightarrow D$}$ \\
  and & & \\
  & $H_{2}:$ & any solution of \eref{eq_app_SY_nonautonomous_DE}
    with initial condition in $D$ is bounded for $\tau \!\geq\! \tau_{0}$,\\
  && for some $\tau_{0}$ sufficiently large.
\end{tabular}

\begin{theorem}
  If $H_{1}$ and $H_{2}$ are satisfied and any solution of
  \eref{eq_app_SY_autonomousDE} with initial condition in $D$ satisfies
  \begin{displaymath}
    \lim_{\tau \rightarrow +\infty} \hat{\vec{x}}(\tau) = \vec{a},
  \end{displaymath}
  then any solution of \eref{eq_app_SY_nonautonomous_DE} with initial
  condition in $D$ satisfies
  \begin{displaymath}
    \lim_{\tau \rightarrow +\infty} \bar{\vec{x}}(\tau) = \vec{a}.
  \end{displaymath} \label{theorem_SY}
\end{theorem} 

\noindent We make use of this theorem in appendix~B.2.


\stepcounter{subsection}
\subsection*{Appendix~\thesubsection.~Limits at late times of
  $(\hat{\Sigma}_{+},\hat{R},\hat{Z})$}

The components of the DE~\eref{eq_limits_autonomous_DE},
$\hat{\vec{x}}' = \vec{f}(\hat{\vec{x}})$, are given by 
\begin{eqnarray}
  \eqalign{
  \hat{\Sigma}'_{+} &= -(2-\hat{Q})\hat{\Sigma}_{+} - \hat{R}^{2} + \hat{Z}^{2}, \\
  \hat{R}' &= (\hat{Q}+\hat{\Sigma}_{+}-1) \hat{R}, \\
  \hat{Z}' &= (\hat{Q}-\hat{\Sigma}_{+}) \hat{Z},}
    \label{eq_app_limits_hatted_DE}
\end{eqnarray}
where
\begin{eqnarray}
  \hat{Q} &= 2 \hat{\Sigma}_{+}^{2} + \hat{R}^{2} + \case{1}{2}(3\gamma-2)
    \hat{\Omega}, \label{eq_app_limits_Q_hat} \\
  \hat{\Omega} &= 1 - \hat{\Sigma}_{+}^{2} - \hat{R}^{2} - \hat{Z}^{2} .
    \label{eq_app_limits_Omega_hat}
\end{eqnarray}
One can also form an auxiliary DE for $\hat{\Omega}$ using
\eref{eq_app_limits_hatted_DE} and \eref{eq_app_limits_Omega_hat} to find that
\begin{equation}
  \hat{\Omega} ' = [ 2\hat{Q} - (3\gamma-2)] \hat{\Omega} .
    \label{eq_app_limits_Omega_DE}
\end{equation}
We consider the state space $S$ of the DE~\eref{eq_app_limits_hatted_DE}
defined by the inequalities
\begin{equation}
  \hat{R} > 0, \quad \hat{Z} > 0, \quad \hat{\Omega} > 0. 
    \label{eq_app_limits_S}
\end{equation}
These inequalities in conjunction with \eref{eq_app_limits_Omega_hat} imply
that the state space $S$ is the interior of one quarter of a sphere.
We also consider the dynamics on the invariant set $S_{\hat{\Omega}}$ defined
by the following restrictions:
\begin{displaymath}
  S_{\hat{\Omega}} : \quad \hat{\Omega} = 0, \quad \hat{R} > 0, \quad
    \hat{Z} > 0.
\end{displaymath}

The DE~\eref{eq_app_limits_hatted_DE} admits a positive monotone function
\begin{equation}
  \chi = \frac{ \hat{\Omega} }{ \hat{R} \hat{Z} }, \label{eq_app_limits_chi}
\end{equation}
which satisfies
\begin{equation}
  \chi ' = 3(1-\gamma) \chi , \label{eq_app_limits_chi_prime}
\end{equation}
on the set $S$. Thus, if $\gamma \not= 1$ there are no equilibrium points,
periodic orbits and homoclinic orbits in $S$ (see WE, proposition~4.2).
It is immediate upon integrating \eref{eq_app_limits_chi_prime} and using the
boundedness of $\hat{R}$ and $\hat{Z}$ that for any $\hat{\vec{x}} \in S$
\begin{equation}
  \omega(\hat{\vec{x}}) \subseteq \bar{S}_{\hat{\Omega}}, \qquad
    \text{if} \quad 1 < \gamma \leq 2 . \label{eq_app_limits_omegaA}
\end{equation}

\begin{figure}
  \begin{indented} \item[] \includegraphics[scale=0.85]{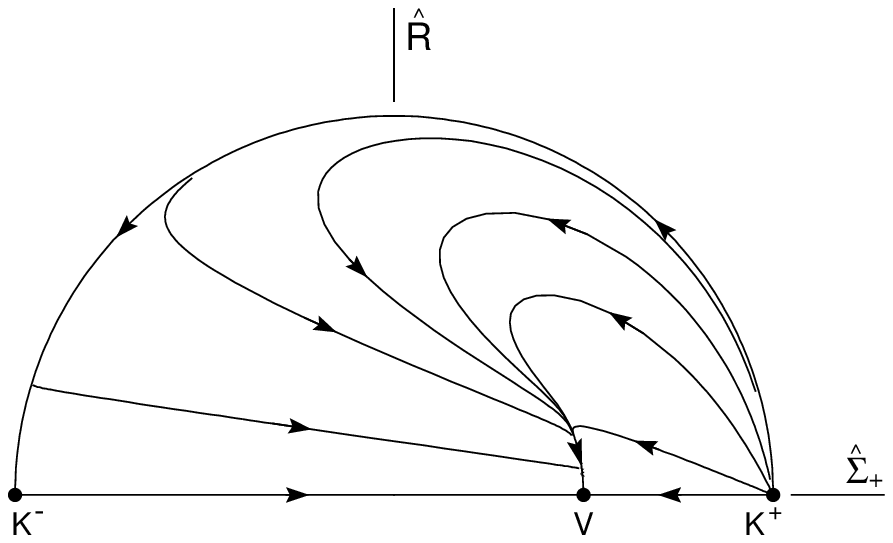}
    \end{indented}
  \caption{\label{fig_Omega0}Orbits in the invariant set $S_{\hat{\Omega}}$.}
\end{figure}

The flow on the invariant set $S_{\hat{\Omega}}$ is depicted in
\fref{fig_Omega0}, which shows the projection of the surface $\hat{\Omega} = 0$
onto the $\hat{\Sigma}_{+} \hat{R}$-plane. The essential features are the
existence of three equilibrium points
\begin{displaymath}
  \eqalign{
    \mathsf{K}^{\pm}: &\quad (\hat{\Sigma}_{+},\hat{R},\hat{Z}) = 
      (\pm 1, 0, 0), \\
    \mathsf{V}: &\quad (\hat{\Sigma}_{+},\hat{R},\hat{Z}) = 
      \left( \case{1}{2}, 0, \case{\sqrt{3}}{2} \right), }
\end{displaymath}
which lie on the boundary of $S_{\hat{\Omega}}$, and the fact that there are
no periodic orbits or flow-connected heteroclinic cycles on $S_{\hat{\Omega}}$.
Thus the only potential $\omega$-limit sets in $S_{\hat{\Omega}}$ are the
equilibrium points $\mathsf{K}^{\pm}$ and $\mathsf{V}$ and hence for any
$\hat{\vec{x}} \in S$, the $\omega$-limit set is one of $\mathsf{K}^{\pm}$
or $\mathsf{V}$. The point $\mathsf{K}^{+}$ can be excluded since it is a
local source in $S$. The point $\mathsf{K}^{-}$ can be excluded by considering
the evolution equation for $\hat{Z}$, which is of the form
\begin{displaymath}
  \hat{Z} ' = z(\hat{\Sigma}_{+},\hat{R},\hat{Z}) \hat{Z}.
\end{displaymath}
Since $z(\mathsf{K}^{-}) = z(-1,0,0) = 3$ and $\hat{Z} = 0$ at
$\mathsf{K}^{-}$, it follows that $\lim_{\tau \rightarrow +\infty} \hat{Z}
\not= 0$ and hence that an orbit in $S$ cannot be future asymptotic to
$\mathsf{K}^{-}$. We thus conclude that $\omega(\hat{\vec{x}}) = \mathsf{V}$
for any $\hat{\vec{x}} \in S$. Equivalently,
\begin{equation}
  \lim_{\tau \rightarrow +\infty} (\hat{\Sigma}_{+},\hat{R},\hat{Z}) =
    \left( \case{1}{2}, 0, \case{\sqrt{3}}{2} \right).
    \label{eq_app_limits_main_result}
\end{equation}

\begin{figure}
  \begin{indented} \item[] \includegraphics[scale=0.6]{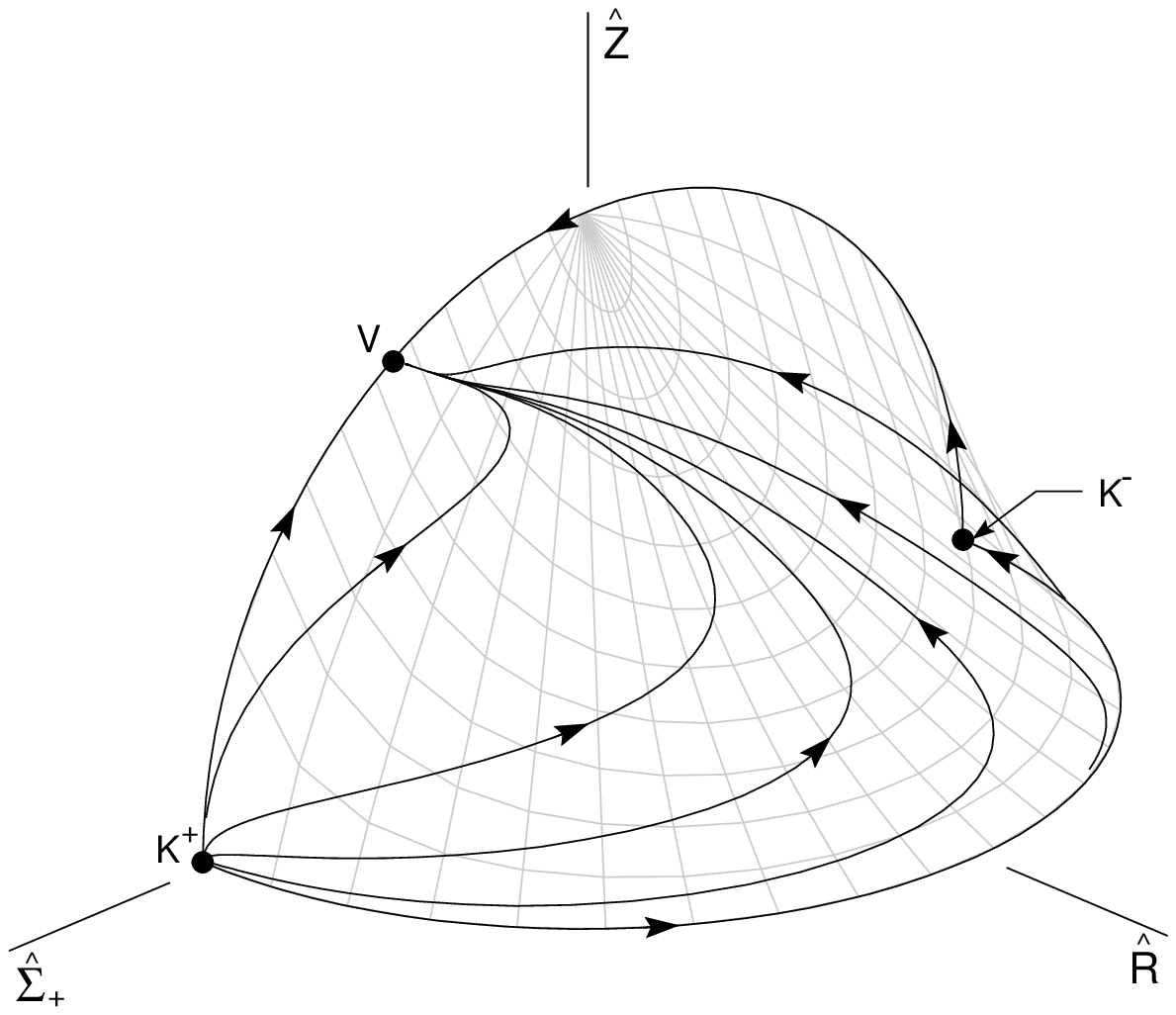}
    \end{indented}
  \caption{\label{fig_gamma1}Orbits in the invariant set
    $\dsty \frac{ \hat{\Omega} }{ \hat{R} \hat{Z} } = 2k$, $\gamma = 1$.}
\end{figure}

Finally, we consider the case $\gamma = 1$. By \eref{eq_app_limits_chi_prime}
the function $\chi$ defined in \eref{eq_app_limits_chi} describes a conserved
quantity
\begin{equation}
  \frac{ \hat{\Omega} }{ \hat{R} \hat{Z} } = 2k,
    \label{eq_app_limits_chi_conserved}
\end{equation}
where $k > 0$ is a constant that depends on the initial conditions.
We see that for all $k > 0$ the surfaces described by
\eref{eq_app_limits_chi_conserved} foliate the state space $S$ and intersect
the vacuum boundary $\hat{\Omega} = 0$ at $\hat{R} = 0$ and $\hat{Z} = 0$
(see~\fref{fig_gamma1}).

We can use similar techniques from before to conclude that the equilibrium
points $\mathsf{K}^{\pm}$ are asymptotically unstable. Since the orbits are
constrained to lie on the two-dimensional family of invariant sets defined by
\eref{eq_app_limits_chi_conserved}, it follows that $\omega(\hat{\vec{x}}) = 
\mathsf{V}$ for any $\hat{\vec{x}} \in S$. Thus any solution of
the DE~\eref{eq_app_limits_hatted_DE} in the set $S$ also satisfies
\eref{eq_app_limits_main_result} if $\gamma = 1$.


\stepcounter{subsection}
\subsection*{Appendix~\thesubsection.~Limits at late times of
  $(\bar{\Sigma}_{+},\bar{R},\bar{Z})$}

We now apply theorem~\ref{theorem_SY} to prove \eref{eq_limits_barred_vector},
namely
\begin{displaymath}
  \lim_{\tau \rightarrow +\infty} \bar{\vec{x}} = \vec{a},
\end{displaymath}
where $\bar{\vec{x}} = (\bar{\Sigma}_{+},\bar{R},\bar{Z})$ and
$\vec{a} = \left(\frac{1}{2},0,\frac{\sqrt{3}}{2}\right)$, considering the
cases $1 < \gamma \leq 2$ and $\gamma = 1$ simultaneously.

We begin by defining the subset $D$ in theorem~\ref{theorem_SY} by
\begin{displaymath}
  \Sigma_{+}^{2} + R^{2} + Z^{2} < 1.
\end{displaymath}
We now verify the hypotheses $H_{1}$ and $H_{2}$. Firstly, let
$\vec{w} : [\tau_{0},+\infty) \rightarrow D$ be any $C^{0}[\tau_{0},\infty)$
function. Since $\lim_{\tau \rightarrow +\infty} M(\tau) = 0$ it follows
immediately that
\begin{displaymath}
  \lim_{\tau \rightarrow +\infty} \vec{g}(\vec{w}(\tau),\tau) =
  \lim_{\tau \rightarrow +\infty} M(\tau) \left( B_{\bar{\Sigma}_{+}},
    \bar{R} B_{\bar{R}}, \bar{Z} B_{\bar{Z}} \right)
    \Big|_{\bar{\vec{x}} = \vec{w}(\tau)} = \mathbf{0},
\end{displaymath}
showing that $H_{1}$ is satisfied. Secondly, $H_{2}$ is satisfied
since the variables $\bar{\Sigma}_{+}$, $\bar{R}$ and $\bar{Z}$ are bounded
for all $\tau \geq \tau_{0}$ with $\tau_{0}$ sufficiently large. Therefore,
since
\begin{displaymath}
  \lim_{\tau \rightarrow +\infty} \hat{\vec{x}}(\tau) = \vec{a}
\end{displaymath}
for all initial conditions $\hat{\vec{x}}(\tau_{0})$ in $D$
(see \eref{eq_app_limits_main_result}), theorem~\ref{theorem_SY} implies that
\begin{equation}
  \lim_{\tau \rightarrow +\infty} \bar{\vec{x}}(\tau) = \vec{a}
    \label{eq_app_limits_SYmainresult}
\end{equation}
for all initial conditions $\bar{\vec{x}}(\tau_{0})$ in $D$.

Finally, we need to show that any initial condition $\vec{x}(\tau_{0}) =
(\Sigma_{+},R,Z) \big|_{\tau=\tau_{0}}$, $M(\tau_{0})$, $\psi(\tau_{0})$
for the DE~\eref{eq_eveqn_full_DE}, subject to $\Omega > 0$ and
\eref{eq_eveqn_restrictionsC}, determines an initial condition
$\bar{\vec{x}}(\tau_{0})$ in $D$ for the DE~\eref{eq_limits_nonautonomous_DE},
so that \eref{eq_app_limits_SYmainresult} is satisfied.

Indeed, since $\lim_{\tau \rightarrow +\infty} \psi = +\infty$, we can without
loss of generality restrict the initial condition $\psi(\tau_{0})$ to be a
multiple of $\pi$. This requirement can be achieved by simply following the
solution determined by the original initial condition until this condition is
satisfied. It follows from this condition, in conjunction with
\eref{eq_limits_cov} and the restriction $\Omega > 0$ applied to
\eref{eq_eveqn_Omega}, that
\begin{displaymath}
  \left. \left( \bar{\Sigma}_{+}^{2} + \bar{R}^{2} + \bar{Z}^{2} \right)
    \right|_{\tau = \tau_{0}} = \left. \left( \Sigma_{+}^{2} + R^{2}
    + Z^{2} \right) \right|_{\tau=\tau_{0}} < 1,
\end{displaymath}
so that $\bar{\vec{x}}(\tau_{0}) \in D$.


\stepcounter{subsection}
\subsection*{Appendix~\thesubsection.~The limit of $M/R$ at late times}

In analogy to \eref{eq_limits_cov}, we define a variable $\bar{M}$ by
\begin{equation}
  \bar{M} = M \left( 1 + \case{1}{4} R^{2} M \sin 2 \psi \right).
    \label{eq_app_MoverR_Mbar}
\end{equation}
It follows from \eref{eq_eveqn_full_DE} that the evolution equation for
$\bar{M}$ is of the form
\begin{equation}
  \bar{M}' = -(\bar{Q} + 2 \bar{\Sigma}_{+} + M B_{\bar{M}} ) \bar{M},
    \label{eq_app_MoverR_M_eveqn}
\end{equation}
where $B_{\bar{M}}$ is a bounded function for $\tau$ sufficiently large.
By using \eref{eq_limits_barred_DE} we obtain
\begin{equation}
  \left( \frac{\bar{M}}{\bar{R}} \right)' = \left( -\case{3}{2} +
    h(\bar{\vec{x}},M,\psi) \right) \left( \frac{\bar{M}}{\bar{R}} \right),
    \label{eq_app_MoverR_MoverR_eveqn}
\end{equation}
where
\begin{displaymath}
  h(\bar{\vec{x}},M,\psi) = \case{5}{2}- 2 \bar{Q} - 3 \bar{\Sigma}_{+} + 
    M B_{*}
\end{displaymath}
and $B_{*}$ is a bounded function for $\tau$ sufficiently large. It follows
from \eref{eq_limits_M}, \eref{eq_limits_cov}, \eref{eq_limits_Q_bar} and
theorem~\ref{theorem_limits} that $\lim_{\tau \rightarrow +\infty}
h(\bar{\vec{x}},M,\psi) = 0$. Consequently, \eref{eq_app_MoverR_MoverR_eveqn}
implies that
\begin{equation}
  \frac{\bar{M}}{\bar{R}} = \bigO \left( \rme^{(-3/2+\delta)\tau} \right),
    \qquad \text{if} \quad 1 \leq \gamma \leq 2,
    \label{eq_app_MoverR_decayrate}
\end{equation}
as $\tau \rightarrow +\infty$ for any $\delta > 0$. Therefore,
$\lim_{\tau \rightarrow +\infty} \bar{M}/\bar{R} = 0$, which implies that
$\lim_{\tau \rightarrow +\infty} M/R = 0$, on account of
\eref{eq_app_MoverR_Mbar} and \eref{eq_limits_cov}.


\stepcounter{section}
\section*{Appendix~\thesection.~Derivation of the asymptotic expansions
  \eref{eq_asympt_state_variables} and \eref{eq_asympt_Omega}}

In this appendix we give details of the derivation of the asymptotic
expansions \eref{eq_asympt_state_variables} and \eref{eq_asympt_Omega}
for $\Sigma_{+}$, $R$, $Z$, $M$ and $\Omega$ as $\tau \rightarrow +\infty$.
As was shown in appendix~B.2, all solutions $\hat{\vec{x}}(\tau)$ of the
DE~\eref{eq_limits_autonomous_DE} in the set $S$ (defined by
\eref{eq_app_limits_S}) are future asymptotic to the equilibrium point
$(\hat{\Sigma}_{+},\hat{R},\hat{Z}) = \left(\frac{1}{2},0,
\frac{\sqrt{3}}{2}\right)$. Since this equilibrium point is non-hyperbolic,
centre manifold theory is required in order to compute the decay rates of
$\hat{\vec{x}}(\tau)$. In addition, the analysis has to be split into two
cases, $1 < \gamma \leq 2$ and $\gamma = 1$, as described in appendices~C.1
and C.2. We refer the reader to Carr (1981, pp~1--13) for an introductory
discussion of centre manifold theory.

The next step is to show that the asymptotic expansions for
$\hat{\vec{x}}(\tau)$ are compatible with the non-autonomous
DE~\eref{eq_limits_nonautonomous_DE}. In order to show this we need an
asymptotic decay rate for $M$, which we obtain as follows.
Equation~\eref{eq_app_MoverR_decayrate} and the boundedness of $\bar{R}$
imply that $\bar{M} = \bigO(\rme^{(-3/2+\delta)\tau})$, as $\tau \rightarrow
+\infty$, for any $\delta > 0$. Using \eref{eq_app_MoverR_Mbar} we can
conclude that
\begin{equation}
  M = \bigO \left(\rme^{(-3/2+\delta)\tau}\right), \label{eq_app_asympt_M}
\end{equation}
as $\tau \rightarrow +\infty$, for any $\delta > 0$. We then write the
non-autonomous DE~\eref{eq_limits_nonautonomous_DE} in integral form
\begin{equation}
  \bar{\vec{x}}(\tau) = \bar{\vec{x}}(\tau_{0}) + \int_{\tau_{0}}^{\tau}
    \left[ \vec{f}(\bar{\vec{x}}(s)) + \vec{g}(\bar{\vec{x}}(s),s) \right]\,
    \rmd s, \label{eq_app_asympt_nonautonomousDE_integralform}
\end{equation}
and can verify that the expansions for $\hat{\vec{x}}(\tau)$ are compatible
with \eref{eq_app_asympt_nonautonomousDE_integralform} using
\eref{eq_limits_g} and \eref{eq_app_asympt_M}.


\stepcounter{subsection}
\subsection*{Appendix~\thesubsection.~$1 < \gamma \leq 2$}

In order to put the DE~\eref{eq_limits_autonomous_DE} in the canonical form
required for application of centre manifold theory, we begin by making
the translation
\begin{equation}
  (\hat{\Sigma}_{+},\hat{R},\hat{Z}) = \left( y+\case{1}{2}, x,
    z+\case{\sqrt{3}}{2} \right). \label{eq_app_asympt_cov_c1}
\end{equation}
We can now apply theorem~3 in Carr (1981) to obtain an approximation to
the local centre manifold through the point $(x,y,z)=(0,0,0)$ given by
\begin{eqnarray}
  \eqalign{
  y &= -x^{2} - 2x^{4} - \case{40}{3} x^{6} + \bigO(x^{8}), \\
  z &= \case{1}{\sqrt{3}} x^{4} + \case{28}{3\sqrt{3}} x^{6} + 
    \bigO(x^{8}) . } \label{eq_app_asympt_cmapprox_c1}
\end{eqnarray}
Upon applying theorem~1 in Carr (1981), we find that the governing DE for
the flow on the centre manifold is given by
\begin{equation}
  x' = -2x^{3} - 4x^{5} -32x^{7} + \bigO(x^{9}).
    \label{eq_app_asympt_cmDE_c1}
\end{equation}
To obtain the asymptotic form of $x(\tau)$ from \eref{eq_app_asympt_cmDE_c1},
we follow a procedure analogous to the one outlined in section~3.1 of
Carr (1981) who considers a related problem. It follows that
\begin{equation}
  x(\tau) = \frac{1}{2\tau} \left[ 1 - \frac{1}{4\tau}( \ln \tau + C_{*} )
    + \bigO(\tau^{-2} \ln^{2} \tau) \right], \label{eq_app_asympt_x_c1}
\end{equation}
as $\tau \rightarrow +\infty$, where $C_{*}$ is a constant that depends
on the initial conditions (which can be set to zero upon appropriate
translation of $\tau$). We have also included the second-order term in
$x(\tau)$ so that the reader can compute the decay rates in
\sref{sec_asympt} to higher order if desired. The decay rates for
$\hat{\Sigma}_{+}$, $\hat{R}$ and $\hat{Z}$ then follow from
\eref{eq_app_asympt_cov_c1}, \eref{eq_app_asympt_cmapprox_c1} and
\eref{eq_app_asympt_x_c1} in conjunction with theorem~2 in Carr (1981).
The compatibility check based on \eref{eq_app_asympt_M} and
\eref{eq_app_asympt_nonautonomousDE_integralform}, in conjunction with the
transformation~\eref{eq_limits_cov}, then lead to the given decay rates for
$\Sigma_{+}$, $R$ and $Z$ in \eref{eq_asympt_state_variables}.

To obtain the asymptotic form of $M$ in \eref{eq_asympt_state_variables},
we substitute the known asymptotic expansions for $\bar{\vec{x}}(\tau)$ into
the evolution equation~\eref{eq_app_MoverR_M_eveqn} for $\bar{M}$, obtaining
\begin{displaymath}
  \bar{M}' = \left[ -\case{3}{2} + \case{3}{4} \tau^{-1}\left( 1 
    + \bigO\left(\case{\ln \tau}{\tau}\right) \right) \right] \bar{M},
\end{displaymath}
on account of \eref{eq_app_asympt_M}. We can now solve this DE and use
\eref{eq_app_MoverR_Mbar} to obtain the decay rate for $M$.

Finally, to obtain the asymptotic expansion for $\Omega$, we define a variable
$\bar{\Omega}$ according to
\begin{equation}
  \bar{\Omega} = \Omega \left( 1 - \case{1}{2} R^{2} M \sin 2\psi \right).
    \label{eq_app_asympt_c1_Omegabar}
\end{equation}
It follows from \eref{eq_eveqn_full_DE} that the evolution equation for
$\bar{\Omega}$ is of the form
\begin{displaymath}
  \bar{\Omega}' = [2\bar{Q} - (3\gamma-2) + M B_{\bar{\Omega}} ] \bar{\Omega},
\end{displaymath}
where $B_{\bar{\Omega}}$ is a bounded function for $\tau$ sufficiently large.
Furthermore, using \eref{eq_limits_barred_DE} we obtain
\begin{displaymath}
  \left( \frac{\bar{\Omega}}{\bar{R}\bar{Z}} \right)' = [-3(\gamma-1) +
    M B_{\ast} ] \frac{\bar{\Omega}}{\bar{R}\bar{Z}} .
\end{displaymath}
We can now solve this DE and then use the decay rates for $\bar{R}$ and
$\bar{Z}$ and \eref{eq_app_asympt_c1_Omegabar} to obtain the asymptotic
expansion for $\Omega$ as stated in \eref{eq_asympt_Omega}.


\stepcounter{subsection}
\subsection*{Appendix~\thesubsection.~$\gamma = 1$}

Unlike the case $1 < \gamma \leq 2$ where the centre manifold is
one-dimensional, when $\gamma = 1$ the centre manifold of the equilibrium
point $(\hat{\Sigma}_{+},\hat{R},\hat{Z}) = \left( \frac{1}{2}, 0,
\frac{\sqrt{3}}{2} \right)$ is two-dimensional. Although the analysis is
potentially more complicated, it can be reduced to a one-dimensional centre
manifold problem by considering the DE~\eref{eq_app_limits_hatted_DE}
on the invariant surfaces $\hat{\Omega}/(\hat{R}\hat{Z}) = 2k$
(see~\eref{eq_app_limits_chi_conserved}), where $k > 0$ is a constant.
Since, the analysis parallels that in appendix~C.1 we only
provide a brief outline of the proof. To begin, we use this first integral
along with \eref{eq_app_limits_Omega_hat} to solve for $\hat{\Sigma}_{+}$
in terms of $\hat{R}$ and $\hat{Z}$ in a neighbourhood around
$\hat{\Sigma}_{+} = \frac{1}{2}$ to obtain
\begin{equation}
  \hat{\Sigma}_{+} = ( 1 - 2k \hat{R}\hat{Z} - \hat{R}^{2} - 
    \hat{Z}^{2} )^{1/2} . \label{eq_app_asympt_Sigma_c2}
\end{equation}
Substituting \eref{eq_app_asympt_Sigma_c2} into the $\hat{R}$ and $\hat{Z}$
evolution equations in \eref{eq_app_limits_hatted_DE} and then making the
transformation
\begin{eqnarray}
  \eqalign{
  \hat{R} &= x , \\
  \hat{Z} &= -\case{k}{2} x + \case{k}{2} y + \case{\sqrt{3}}{2} , }
    \label{eq_app_asympt_cov_c2}
\end{eqnarray}
puts the system into canonical form. We find that the centre manifold through
the point $(x,y)=(0,0)$ is approximated by
\begin{equation}
  y = \case{2}{3} x^{3} + \case{2\sqrt{3}}{9k}(3+10k^{2}) x^{4} +
    \case{22}{3} (2+k^{2}) x^{5} + \bigO(x^{6}),
    \label{eq_app_asympt_cmapprox_c2}
\end{equation}
the corresponding DE on the centre manifold is
\begin{equation}
  x' = -\sqrt{3}\, k x^{2} + (k^{2}-2)\, x^{3} - 2\sqrt{3}\, k x^{4} +
    \bigO(x^{5}) , \label{eq_app_asympt_cmDE_c2}
\end{equation}
and the asymptotic expansion for $x(\tau)$ is given by
\begin{equation}
  x(\tau) = \frac{C_{R}}{\tau}\left[1 + \frac{1}{3\tau}\left( (1-6C_{R}^{2})
    \ln \tau + C_{*} \right) + \bigO( \tau^{-2} \ln^{2} \tau ) \right ],
    \label{eq_app_asympt_x_c2}
\end{equation}
as $\tau \rightarrow +\infty$, where $C_{*}$ is a constant that depends
on the initial conditions. We note that the constant $C_{R}$ is related to the
constant $k$ through $C_{R} = 1/(\sqrt{3}\,k)$.

The asymptotic decay rates for $\hat{R}$ and $\hat{Z}$ follow from
\eref{eq_app_asympt_cov_c2}, \eref{eq_app_asympt_cmapprox_c2} and
\eref{eq_app_asympt_x_c2}, in conjunction with theorem~2 in Carr (1981).
The expansion for $\hat{\Sigma}_{+}$ is obtained from
\eref{eq_app_asympt_Sigma_c2}. Dropping the hats as in appendix~C.1 yields the
decay rates for $\Sigma_{+}$, $R$ and $Z$ as stated in
equation~\eref{eq_asympt_state_variables}. The asymptotic form of
$M$ is obtained in an analogous fashion to that outlined in appendix~C.1.
Finally, the asymptotic expansion for $\Omega$ as stated in
\eref{eq_asympt_Omega} is obtained directly from \eref{eq_eveqn_Omega}.


\stepcounter{section}
\section*{Appendix~\thesection.~The Weyl curvature tensor}

In this appendix we give an expression for the Weyl curvature parameter
$\mathcal{W}$ in terms of the Hubble-normalized variables $\Sigma_{+}$,
$R$, $Z$, $M$ and $\psi$. Let $E_{\alpha\beta}$ and $H_{\alpha\beta}$ be the
components of the electric and magnetic parts of the Weyl tensor relative to
the group invariant frame with $\vec{e}_{0} = \vec{u}$. Since
$E_{\alpha\beta}$ and $H_{\alpha\beta}$ are diagonal and trace-free (see WE,
chapter 6, appendix), they each have two independent components. In analogy
with \eref{eq_eveqn_Npm} we define
\begin{eqnarray}
  \eqalign{
  \mathcal{E}_{+} = \case{1}{2}(\mathcal{E}_{22}+\mathcal{E}_{33}), &\qquad
  \mathcal{E}_{-} = \case{1}{2\sqrt{3}}(\mathcal{E}_{22}-\mathcal{E}_{33}), \\
  \mathcal{H}_{+} = \case{1}{2}(\mathcal{H}_{22}+\mathcal{H}_{33}), &\qquad
  \mathcal{H}_{-} = \case{1}{2\sqrt{3}}(\mathcal{H}_{22}-\mathcal{H}_{33}), }
    \label{eq_app_Weyl_a}
\end{eqnarray}
where $\mathcal{E}_{\alpha\beta}$ and $\mathcal{H}_{\alpha\beta}$ are the
dimensionless counterparts of $E_{\alpha\beta}$ and $H_{\alpha\beta}$,
defined by
\begin{equation}
  \mathcal{E}_{\alpha\beta} = \frac{E_{\alpha\beta}}{H^{2}}, \qquad
  \mathcal{H}_{\alpha\beta} = \frac{H_{\alpha\beta}}{H^{2}}.
    \label{eq_app_Weyl_b}
\end{equation}
It follows from \eref{eq_limits_Weyl_defn}, \eref{eq_app_Weyl_a} and
\eref{eq_app_Weyl_b} that
\begin{equation}
  \mathcal{W}^{2} = \mathcal{E}_{+}^{2} + \mathcal{E}_{-}^{2}
    + \mathcal{H}_{+}^{2} + \mathcal{H}_{-}^{2}. \label{eq_app_Weyl_c}
\end{equation}
The desired expressions for $\mathcal{E}_{\pm}$ and $\mathcal{H}_{\pm}$ can be
obtained directly from equations (6.36) and (6.37) in WE, using
\eref{eq_eveqn_cov} in the present paper:
\begin{eqnarray}
  \eqalign{
  \mathcal{E}_{+} &= \Sigma_{+}(1+\Sigma_{+}) + \case{1}{2}
    R^{2}(1-3 \cos 2 \psi) - Z^{2} - 3 M^{2} Z^{4} , \\
  \mathcal{H}_{+} &= -\case{3}{2} R^{2} \sin 2 \psi + \case{9}{2} M
    \Sigma_{+} Z^{2} , \\
  \mathcal{E}_{-} &= \frac{2R}{M} \left[ \sin \psi + \case{1}{2}M( 1 -
    2 \Sigma_{+}) \cos \psi + \case{3}{2} M^{2} Z^{2} \sin \psi \right] , \\
  \mathcal{H}_{-} &= \frac{2R}{M} \left[ -\cos \psi - \case{3}{2} M \Sigma_{+}
    \sin \psi - \case{3}{4} M^{2} Z^{2} \cos \psi \right] . }
    \label{eq_app_Weyl_explicit}
\end{eqnarray}


\section*{References}

\begin{harvard}

\item[] Barrow J D and Gaspar Y 2001 Bianchi VIII empty futures \CQG
  \textbf{18} 1809--22

\item[] Barrow J D and Hervik S 2002 The Weyl tensor in spatially homogeneous
  cosmological models \textit{Preprint} gr-qc/0206061

\item[] Berger B K, Isenberg J and Weaver M 2001 Oscillatory approach to the
  singularity in vacuum spacetimes with $T^{2}$ isometry
  \textit{Phys.\ Rev.\ D} \textbf{64} 084006

\item[] Berger B K and Moncrief V 1998 Evidence for an oscillatory singularity
  in generic U(1) symmetric cosmologies on $T^{3} \times R$
  \textit{Phys.\ Rev.\ D} \textbf{58} 064023

\item[] Carr J 1981 \textit{Applications of Centre Manifold Theory}
  (Berlin: Springer)

\item[] Doroshkevich A G, Lukash V N and Novikov I D 1973 The isotropization
  of homogeneous cosmological models \textit{Sov.\ Phys.-JETP}
  \textbf{37} 739--46

\item[] Ellis G F R and MacCallum M A H 1969 A class of homogeneous
  cosmological models \textit{Commun.\ Math.\ Phys.}\ \textbf{12} 108--41

\item[] Hewitt C G, Bridson R and Wainwright J 2001 The asymptotic regimes
  of tilted Bianchi II cosmologies \textit{Gen.\ Rel.\ Grav.}\
  \textbf{33} 65--94

\item[] Hewitt C G, Horwood J T and Wainwright J 2002 Asymptotic dynamics of
  the exceptional Bianchi cosmologies (submitted to \CQG)

\item[] Hewitt C G and Wainwright J 1992 Dynamical systems approach to tilted
  Bianchi cosmologies: irrotational models of type V \textit{Phys.\ Rev.\ D}
  \textbf{46} 4242--52

\item[] Lukash V N 1974 Some peculiarities in the evolution of homogeneous
  anisotropic cosmological models \textit{Sov.\ Astron.}\ \textbf{18}
  164--9

\item[] Nilsson U S, Hancock M J and Wainwright J 2000 Non-tilted Bianchi
  VII$_{0}$ models---the radiation fluid \CQG \textbf{17} 3119--34

\item[] Ringstr\"{o}m H 2000 The Bianchi IX attractor \textit{Ann.\ H.\
  Poincar\'{e}} \textbf{2} 405--500

\item[] Ringstr\"{o}m H 2001 The future asymptotics of Bianchi VIII vacuum
  solutions \CQG \textbf{18} 3791--823

\item[] Strauss A and Yorke J A 1967 On asymptotically autonomous differential
  equations \textit{Mathematical Systems Theory} \textbf{1} 175--82

\item[] Wainwright J 2000 Asymptotic self-similarity breaking in cosmology
  \textit{Gen.\ Rel.\ Grav.}\ \textbf{32} 1041--54 

\item[] Wainwright J and Ellis G F R (eds) 1997 \textit{Dynamical Systems in
  Cosmology} (Cambridge: Cambridge University Press)

\item[] Wainwright J, Hancock M J and Uggla C 1999 Asymptotic self-similarity
  breaking at late times in cosmology \CQG \textbf{16} 2577--98

\item[] Wainwright J and Hsu L 1989 A dynamical systems approach to Bianchi
  cosmologies: orthogonal models of class A \CQG \textbf{6} 1409--31

\item[] Wald R M 1983 Asymptotic behaviour of homogeneous cosmological models
  in the presence of a positive cosmological constant \textit{Phys.\ Rev.\
  D} \textbf{28} 2118--20

\item[] Weaver M, Isenberg J and Berger B K 1998 Mixmaster behaviour in
  inhomogeneous cosmological spacetimes \textit{Phys.\ Rev.\ Lett.}\
  \textbf{80} 2984--7

\end{harvard}

\end{document}